\documentclass[envcountsame]{llncs}

\usepackage{amssymb}
\title{An Effective Extension of the Wagner \\ Hierarchy to Blind Counter Automata}
\author{Olivier Finkel\inst{}}
\institute{Equipe de Logique Math\'ematique \\
 U.F.R. de Math\'ematiques, Universit\'e Paris 7 \\ 2 Place Jussieu 75251 Paris
 cedex 05, France \\ \email{finkel@logique.jussieu.fr}.}

\date{}

\begin{document}

\spnewtheorem{Rem}[theorem]{Remark}{\bfseries}{\itshape}
\spnewtheorem{Exa}[theorem]{Example}{\bfseries}{\itshape}

\newcommand{\fa}{\forall}
\newcommand{\Ga}{\Gamma}
\newcommand{\Gas}{\Gamma^\star}
\newcommand{\Si}{\Sigma}
\newcommand{\Sis}{\Sigma^\star}
\newcommand{\Sio}{\Sigma^\om}
\newcommand{\ra}{\rightarrow}
\newcommand{\hs}{\hspace{12mm}

\noi}
\newcommand{\lra}{\leftrightarrow}
\newcommand{\la}{language}
\newcommand{\ite}{\item}
\newcommand{\abs}{\{a, b\}^\star}
\newcommand{\abcs}{\{a, b, c \}^\star}
\newcommand{\ol}{ $\omega$-language}
\newcommand{\orl}{ $\omega$-regular language}
\newcommand{\om}{\omega}
\newcommand{\nl}{\newline}
\newcommand{\noi}{\noindent}
\newcommand{\tla}{\twoheadleftarrow}
\newcommand{\de}{deterministic }
\newcommand{\proo}{\noi {\bf Proof.} }
\newcommand{\aut}{{\bf $\mathcal{A}$ }}

\maketitle

\begin{abstract}
\noi The extension of the Wagner hierarchy to blind counter automata accepting infinite words 
with a Muller acceptance condition is effective. We determine precisely this hierarchy. 
\end{abstract}

\noi {\small {\bf Keywords:} \ol s; blind counter automata; 
effective extension of the  Wagner hierarchy; topological properties;
Wadge hierarchy; Wadge games.}

\section{Introduction}

Regular \ol s are accepted by (\de) Muller automata. Finite machines having a 
stronger expressive power when reading infinite words have also been investigated \cite{sta}. 
Recently Engelfriet and Hoogeboom studied {\bf X}-automata, i.e. automata 
equipped with a storage type {\bf X}, including the cases of pushdown automata, Turing 
machines, Petri nets \cite{eh}. A way to investigate the expressive power of such machines is 
to study the topological complexity of the \ol s they accept. For \de machines, it is shown in 
\cite{eh} that every {\bf X}-automaton accepts boolean combinations of ${\bf \Pi^0_2 }$-sets. 
Hence in order to distinguish the different storage types it turned out  that the study of 
the Wadge hierarchy is suitable. The Wadge hierarchy is a great refinement of the Borel 
hierarchy, recently studied by Duparc \cite{dup}. The Wadge hierarchy of \orl s has been 
determined in an efective way by Wagner \cite{wag}. 
 Several extensions of this hierarchy have been recently 
determined as the extension to \de pushdown automata, to $k$-blind counter automata, 
\cite{dfr} \cite{dupcf} \cite{wadpn}. 
We present here the extension to (one) blind counter automata, which is the first 
known {\bf effective} extension. We  study  Muller blind counter automata (MBCA), and define 
chains and superchains as Wagner did for Muller automata. The essential difference between 
the two hierarchies relies on the existence of superchains of transfinite 
 length $\alpha < \om^2$
for MBCA. The hierarchy is effective and leads to effective winning strategies in Wadge 
games between MBCA. The hierarchy of Muller automata equipped with several 
blind counters is presented in a non effective way in \cite{wadpn}\cite{dfr}.

\section{Regular and  Blind Counter \ol s}

We assume the reader to be familiar with the theory of formal \la s and 
of \orl s, see for example \cite{hu69} ,\cite{tho}.
We first recall some  definitions and results concerning $\om$-regular 
\la s and omega pushdown automata and introduce blind counter automata as a 
special case of pushdown automata  \cite{tho} \cite{sta}.
\nl
When $\Si$ is a finite alphabet, a finite string (word) over $\Si$ is any 
sequence $x=x_1\ldots x_k$ , where $x_i\in\Sigma$ 
for $i=1,\ldots ,k$ ,and  $k$ is an integer $\geq 1$. The length
 of $x$ is $k$, denoted by $|x|$ .
 If  $|x|=0$ , $x$ is the empty word denoted by $\lambda$. 
\nl we write $x(i)=x_i$  and $x[i]=x(1)\ldots x(i)$ for $i\leq k$ and $x[0]=\lambda$.
 $\Sis$  is the set of finite words over $\Sigma$.
The first infinite ordinal is $\om$.
 An $\om$-word over $\Si$ is an $\om$ -sequence $a_1\ldots a_n\ldots  $, where 
$a_i \in\Sigma , \fa i\geq 1$.
 When $\sigma$ is an $\om$-word over $\Si$, we write
 $\sigma =\sigma(1)\sigma(2)\ldots \sigma(n) \ldots $
 and $\sigma[n]=\sigma(1)\sigma(2)\ldots \sigma(n)$ 
the finite word of length n, prefix of $\sigma$.
 The set of $\om$-words over  the alphabet $\Si$ is denoted by $\Si^\om$.
 An  $\om$-language over an alphabet $\Sigma$ is a subset of  $\Si^\om$.

\hs The usual concatenation product of two finite words $u$ and $v$ is 
denoted $u.v$ (and sometimes just $uv$). This product is extended to the product $u.v$ of a 
finite word $u$ and an $\om$-word $v$.

\hs For $V\subseteq \Sis$, $V^\om = \{ \sigma =u_1\ldots u_n \ldots \in \Si^\om /  u_i\in V, \fa i\geq 1 \}$
is the $\om$-power of $V$.

\hs   R. Mc Naughton established that the expressive
 power of \de Muller automata (DMA) is equal to the expressive power of non \de 
Muller automata (MA) \cite{tho}. An \ol~ is regular iff it is accepted by a Muller automaton. 
The class $REG_\om$ of \orl s is  the $\om$-Kleene closure of the class $REG$ 
of (finitary) regular languages where the $\om$-Kleene closure of a 
family L of  finitary \la s is : 
$$\om-KC(L) = \{ \cup_{i=1}^n U_i.V_i^\om  /  U_i, V_i \in L , \fa i\in [1, n] \}$$

\noi We now define the (blind) one counter machines which we assume here 
to be realtime and deterministic, and the corresponding classes 
of  blind counter \ol s.

\begin{definition}
A (realtime deterministic) pushdown machine (PDM) is a 6-tuple 
$M=(K,\Si,\Ga, \delta, q_0, Z_0)$, where $K$ 
is a finite set of states, $\Sigma$ is a finite input alphabet, 
$\Gamma$ is the finite pushdown alphabet,  
 $q_0\in K$ is the initial state, $Z_0 \in\Ga$ is the start symbol, 
and $\delta$ is a mapping from $K \times \Si \times \Ga $ into 
$K\times \Gas$ . 
\nl
If  $\gamma\in\Ga^{+}$ describes the pushdown store content, 
the leftmost symbol will be assumed to be on " top" of the store.
A configuration of a PDM is a pair $(q, \gamma)$ where $q\in K$ and  $\gamma\in\Gas$.
\nl For $a\in \Si$, $\gamma,\beta\in\Ga^{\star}$
and $Z\in\Ga$, if $(p,\beta)$ is in $\delta(q,a,Z)$, then we write
$a: (q,Z\gamma)\mapsto_M (p,\beta\gamma)$.\nl
$\mapsto_M^\star$ is the transitive and reflexive closure of $\mapsto_M$.
(The subscript $M$ will be omitted whenever the meaning remains clear).
\nl
Let $\sigma =a_1a_2\ldots a_n\ldots$ be an  $\om$-word over $\Si$. 
An infinite sequence of configurations $r=(q_i,\gamma_i)_{i\geq1}$ is called 
a  run of $M$ on $\sigma$, starting in configuration $(p,\gamma)$, iff:
\begin{enumerate}
\ite $(q_1,\gamma_1)=(p,\gamma)$
\ite  for each $i\geq 1$,  $a_i: (q_i,\gamma_i)\mapsto_M(q_{i+1},\gamma_{i+1} )$
\end{enumerate}

\hs For every such run, $In(r)$ is the set of all states entered infinitely
 often during run $r$.
\nl
A  run $r$ of $M$ on $\sigma$ , starting in configuration $(q_0,Z_0)$,
 will be simply called " a run of $M$ on $\sigma$ ".

\hs A one counter machine is a PDM such that $\Ga=\{Z_0, I\}$ where $Z_0$ is the 
bottom symbol and always remains at the bottom of the store. So the pushdown store is used like 
a counter whose value is the integer $n$ if the content of the pushdown store is $I^nZ_0$.
\nl A one blind counter machine is a one counter machine such that every transition 
which is enabled at zero level is also enabled at non zero level, i.e. if 
$\delta(q, a, Z_0)=(p, I^nZ_0)$, for some $p, q\in K$, $a\in \Si$ and $n\geq 0$, then 
$\delta(q, a, I)=(p, I^{n+1})$.  But the converse may not be true, i.e. 
some transition may be enabled at non zero level but not at zero level. 
\end{definition}

\begin{definition} A Muller (realtime \de)  blind counter automaton ({\bf MBCA}) is a 7-tuple
 ${\bf \mathcal{A}}=(K,\Si,\Gamma, \delta, q_0, Z_0, \mathcal{F})$ where 
${\bf \mathcal{A'}}=(K,\Si,\Gamma, \delta, q_0, Z_0)$
is a (realtime \de)  one blind counter machine and $\mathcal{F}\subseteq 2^K$ 
is the collection of designated state sets.
\nl
The \ol~ accepted by $M$ is 
$L({\bf \mathcal{A}})= \{  \sigma\in\Si^\om$ / there exists a  run r
 of \aut on $\sigma$ such that $In(r) \in \mathcal{F} \}$.
\nl The class of \ol s accepted by {\bf MBCA} will be denoted  {\bf BC}.  
\end{definition}

\begin{Rem}
Machines we call here one  blind counter machines are sometimes called one partially 
 blind counter machines as in \cite{gre}.
\end{Rem}

\begin{Rem}
If $M$ is a \de pushdown machine , then for every $\sigma\in\Si^\om$,
 there exists at most one run $r$ of $M$ on $\sigma$ determined by the starting configuration.
  Each \ol~ accepted by a Muller \de pushdown automaton 
($DMPDA$)  can be accepted by a $DMPDA$ 
such that for every $\sigma\in\Si^\om$, there exists such  a 
  run of $M$ on $\sigma$. 
\nl But this is not true for {\bf MBCA} because some words $x$ may be rejected by an  MBCA  
\aut because the machine \aut blocks at zero level when reading $x$. This is connected with 
the fact that the class  {\bf BC} is not closed under complementation 
as it is shown by the following example. 
\end{Rem}

\begin{Exa}
It is easy to see that the \ol~ $L=\{a^nb^pc^\om~/~ p\leq n \}$ is accepted 
by a \de MBCA, but its complement is not accepted by any \de MBCA because 
$L'=\{a^nb^pc^\om~/~ p > n \}$ is not accepted by any \de MBCA. 
\end{Exa}

\section{Topology}

\noi We assume the reader to be familiar with basic notions of topology which
may be found in \cite{ku}\cite{lt} \cite{sta} \cite{pp}.

\hs
Topology is an important tool for the study of \ol s, and leads 
to characterization of several classes of \ol s.
\nl For a finite alphabet $X$, we consider $X^\om$ 
as a topological space with the Cantor topology (see \cite{lt} \cite{sta} \cite{pp}).
 The open sets of $X^\om$ are the sets in the form $W.X^\om$, where $W\subseteq X^\star$.
A set $L\subseteq X^\om$ is a closed set iff its complement $X^\om - L$ is an open set.
The class of open sets of $X^\om$ will be denoted by ${\bf G}$ or by ${\bf \Si^0_1 }$. 
The class of closed sets will be denoted by ${\bf F}$ or by ${\bf \Pi^0_1 }$. 
Closed sets are characterized by the following:

\begin{proposition}
A set $L\subseteq X^\om$ is a closed set of $X^\om$ iff for every $\sigma\in X^\om$, 

$[\fa n\geq 1,  \exists u\in X^\om$  such that $\sigma (1)\ldots \sigma (n).u \in L]$
 implies that $\sigma\in L$.
\end{proposition}

\noi Define now the next classes of the  Hierarchy of Borel sets of finite rank:

\begin{definition}
The classes ${\bf \Si_n^0}$ and ${\bf \Pi_n^0 }$ of the Borel Hierarchy
 on the topological space $X^\om$  are defined as follows:
\nl ${\bf \Si^0_1 }$ is the class of open sets of $X^\om$.
\nl ${\bf \Pi^0_1 }$ is the class of closed sets of $X^\om$.
\nl ${\bf \Pi^0_2 }$  or ${\bf G_\delta }$ is the class of countable intersections of 
 open sets of $X^\om$.
\nl  ${\bf \Si^0_2 }$  or ${\bf F_\sigma }$ is the class of countable unions  of 
closed sets of $X^\om$.
\nl And for any integer $n\geq 1$:
\nl ${\bf \Si^0_{n+1} }$   is the class of countable unions 
of ${\bf \Pi^0_n }$-subsets of  $X^\om$.
\nl ${\bf \Pi^0_{n+1} }$ is the class of countable intersections of 
${\bf \Si^0_n}$-subsets of $X^\om$.
\end{definition}

\noi There is a nice characterization of ${\bf \Pi^0_2 }$-subsets of $X^\om$. 
First define the notion of $W^\delta$:

\begin{definition}
For $W\subseteq X^\star$,
 let: 
\nl $W^\delta=\{\sigma\in X^\om / \exists^\om i$ such that $\sigma[i]\in W\}$.
\nl ($\sigma \in W^\delta$ iff $\sigma$ has infinitely many prefixes in $W$).
\end{definition}

\noi Then we can state the following Proposition:

\begin{proposition}
A subset $L$ of $X^\om$ is a ${\bf \Pi^0_2 }$-subset of $X^\om$ iff there exists 
a set $W\subseteq X^\star$ such that $L=W^\delta$.
\end{proposition}

\noi Mc Naughton's Theorem implies that every \orl~ is a boolean combination 
of $G_\delta$-sets, hence a ${\bf \Delta^0_3 }=({\bf \Pi^0_3 } \cap {\bf \Si^0_3 })$-set. 
This result holds in fact for every \ol~ accepted by a \de {\bf X}-automaton in the sense 
of \cite{eh}, i.e. an automaton equipped with a storage type {\bf X}, including the case of 
the Turing machine. A way to distinguish the expressive power of finite machines reading 
$\om$-words is the 
Wadge hierarchy which we now introduce.

\begin{definition}
For $E\subseteq X^\om$ and $F\subseteq Y^\om$, $E$ is said to be Wadge reducible to $F$
($E\leq _W F)$ iff there exists a continuous function $f: X^\om \ra Y^\om$, such that
$E=f^{-1}(F)$.
\nl $E$ and $F$ are Wadge equivalent iff $E\leq _W F$  and $F\leq _W E$.  
This will be denoted by $E\equiv_W F$. And we shall say that 
$E<_W F$ iff $E\leq _W F$ but not $F\leq _W E$.
\nl  A set $E\subseteq X^\om$ is said to be self dual iff  $E\equiv_W (X^\om-E)$, and otherwise 
it is said to be non self dual.

\end{definition}

\noi
 The relation $\leq _W $  is reflexive and transitive,
 and $\equiv_W $ is an equivalence relation.
\nl The equivalence classes of $\equiv_W $ are called wadge degrees. 
\nl $WH$ is the class of Borel subsets of finite rank of a set  $X^\om$, where  
$X$ is a finite set,
 equipped with $\leq _W $ and with $\equiv_W $.
\nl  For $E\subseteq X^\om$ and $F\subseteq Y^\om$, if   
$E\leq _W F$ and $E=f^{-1}(F)$  where $f$ is a continuous 
function from $ X^\om$  into $Y^\om$, then $f$ is called a continuous reduction of $E$ to 
$F$. Intuitively it means that $E$ is less complicated than $F$ because 
to check whether $x\in E$ it suffices to check whether $f(x)\in F$ where $f$ 
is a continuous function. Hence the Wadge degree of an \ol~
is a measure 
of its topological complexity.

\begin{Rem}
In the above definition, we consider that a subset $E\subseteq  X^\om$ is given
together with the alphabet $X$. This is necessary as it is shown by the following 
example. 
\nl Let $E=\{0, 1\}^\om$ considered as an \ol~ over the alphabet $X=\{0, 1\}$ and let 
$F=\{0, 1\}^\om$ be the same \ol~ considered as an \ol~ over the alphabet $Y=\{0, 1, 2\}$.  
Then $E$ is an open and closed subset of $\{0, 1\}^\om$ but $F$ is a closed and non open 
subset of $\{0, 1, 2\}^\om$. It is easy to check that $E <_W F$ hence $E$ and $F$ are not 
Wadge equivalent.  
\end{Rem}

\noi Then we can define the Wadge class of a set $F$:
\begin{definition}
Let $F$ be a subset of $X^\om$. The wadge class of $F$ is $[F]$ defined by:
$[F]= \{ E / E\subseteq Y^\om$ for a finite alphabet $Y$ and $E\leq _W F \}$. 
\end{definition}

\noi Recall that each Borel class ${\bf \Si^0_n}$ and ${\bf \Pi^0_n}$ is a Wadge class.

\hs  There is a close relationship between Wadge reducibility
 and games which we now introduce. Define first the Wadge game $W(A, B)$ for 
$A\subseteq X_A^\om$ and $B\subseteq X_B^\om$:

\begin{definition} 
The Wadge game  $W(A, B)$ is a game with perfect information between two players,
player 1 who is in charge of $A$ and player 2 who is in charge of $B$.
\nl Player 1 first writes a letter $a_1\in X_A$, then player 2 writes a letter
$b_1\in X_B$, then player 1 writes a letter $a_2\in  X_A$, and so on  \ldots 
\nl The two players alternatively write letters $a_n$ of $X_A$ for player 1 and $b_n$ of $X_B$
for player 2.
\nl After $\om$ steps, the player 1 has written an $\om$-word $a\in X_A^\om$ and the player 2
has written an $\om$-word $b\in X_B^\om$.
\nl The player 2 is allowed to skip, even infinitely often, provided he really write an
$\om$-word in  $\om$ steps.
\nl The player 2 wins the play iff [$a\in A \lra b\in B$], i.e. iff 
\nl  [($a\in A ~{\rm and} ~ b\in B$)~ {\rm or} ~ 
($a\notin A ~{\rm and}~ b\notin B~{\rm and} ~ b~{\rm is~infinite}  $)].
\end{definition}

\noi
Recall that a strategy for player 1 is a function 
$\sigma :(X_B\cup \{s\})^\star\ra X_A$.
And a strategy for player 2 is a function $f:X_A^+\ra X_B\cup\{ s\}$.
\nl $\sigma$ is a winning stategy (w.s.) for player 1 iff he always wins a play when
 he uses the strategy $\sigma$, i.e. when the  $n^{th}$  letter he writes is given
by $a_n=\sigma (b_1\ldots b_{n-1})$, where $b_i$ is the letter written by player 2 
at step $i$ and $b_i=s$ if player 2 skips at step $i$.
\nl A winning strategy for player 2 is defined in a similar manner.

\hs   Martin's Theorem states that every Gale-Stewart Game $G(X)$ (see  \cite{tho} \cite{pp} 
 for more details),  with $X$ a borel set, 
is determined and this implies the following :

\begin{theorem} [Wadge] Let $A\subseteq X_A^\om$ and $B\subseteq X_B^\om$ be two Borel sets, where
$X_A$ and $X_B$ are finite alphabets. Then the Wadge game $W(A, B)$ is determined:
one of the two players has a winning strategy. And $A\leq_W B$ iff the player 2 has a 
winning strategy  in the game $W(A, B)$.
\end{theorem}

\noi Recall that a set $X$ is well ordered by a binary relation $<$ iff $<$ 
 is a linear order on $X$ and there is not any strictly decreasing (for $<$) 
infinite sequence of elements in $X$.

\begin{theorem} [Wadge]
Up to the complement and $\equiv _W$, the class of Borel subsets of finite rank of $X^\om$,
 for $X$ a finite alphabet, is a well ordered hierarchy.
 There is an ordinal $|WH|$, called the length of the hierarchy, and a map
$d_W^0$ from $WH$ onto $|WH|$, such that for all $A, B\in WH$:
\nl $d_W^0 A < d_W^0 B \lra A<_W B $  and 
\nl $d_W^0 A = d_W^0 B \lra [ A\equiv_W B $ or $A\equiv_W B^-]$.
\end{theorem}

\begin{Rem} We do not give here the ordinal $|WH|$. Details may be found in \cite{dup}.
\end{Rem}

\section{Wagner Hierarchy and its Extension to Blind Counter Automata}

\noi Consider now \orl s. Landweber studied first the topological properties of \orl s.
 He characterized the \orl s in each of the Borel classes 
 ${\bf F, G, F_\sigma, G_\delta }$, and showed that one can decide, for an effectively given
\orl~ $L$, whether $L$ is in  ${\bf F, G, F_\sigma}$, or  ${\bf G_\delta }$.
\nl It turned out that an \orl~ is in the class ${\bf G_\delta }$  iff it 
is accepted by a \de B\"uchi automaton.  These results were refined by K. Wagner 
who studied the Wadge Hierarchy of \orl s. 
In fact there is an effective version of the Wadge Hierarchy restricted to \orl s:

\begin{theorem}[Corollary of B\"uchi-Landweber's Theorem \cite{bl}] 
For $A$ and $B$ some $\om$-regular sets,
 one can effectively decide which player has a w.s. in the game $W(A, B)$ and the 
winner has a w.s. given by a transducer.
\end{theorem}

\noi The hierarchy obtained on \orl s is now called the Wagner hierarchy and has 
length $\om^\om$.
Wagner \cite{wag} gave an automata structure characterization, based on notion of chain
 and superchain, for an automaton to be in a given class and showed that 
the Wadge degree of an \orl~ is computable.  Wilke and Yoo  proved  in \cite{wy} 
that this can be done in polynomial time. Wagner's hierarchy 
  has been recently studied by Carton and Perrin 
in connection with the theory of $\om$-semigroups \cite{cp1} 
\cite{cp2}  \cite{pp} and by Selivanov in \cite{se}.

\hs We present in this paper an extension of the Wagner hierarchy to the class of 
blind counter \ol s, using analogous notions of chains and superchains. 
We shall first define  positive and negative loops, next  chains and 
superchains. 
A  crucial fact  which allows this definition 
is the following lemma:

\begin{lemma}\label{loop-lemma}
Let \aut$=(K,\Si,\Gamma, \delta, q_0, Z_0, \mathcal{F})$ be a   
MBCA and $x\in \Sio$ such that there exists an infinite run 
$r=(q_i, I^{n_i}Z_0)_{i\geq 1}$ of \aut over $x$ such that $Inf(r)=F \subseteq K$. Then there 
exist infinitely many integers $i$ such that for all $j\geq i$, $n_j\geq n_i$. Among these 
integers there exist infinitely many integers $i_k$, $k\geq 1$, and a state $q\in K$ 
 such that for all $k\geq 1$, $q_{i_k}=q$. Then there exist two integers $s, s'$ such that 
  between steps $i_s$ and $i_{s'}$ of the run $r$, \aut enters in every state of 
$F$ and in not any other state of $K$, because $Inf(r)=F$. 
\end{lemma}

\proo 
With the hypotheses of the lemma, assume that  $r=(q_i, I^{n_i}Z_0)_{i\geq 1}$ 
is an infinite run of $M$ over $x$. If there exist only 
finitely many integers $i$ such that for all $j\geq i$, $n_j\geq n_i$, then there exists 
a largest one $l$. But then if $j_0$ is an integer $>l$ there exists an integer 
$j_1>j_0$ such that $n_{j_1}<n_{j_0}$. By induction one could construct a sequence 
of integers $(j_k)_{k\geq 0}$ such that for all $k$, $n_{j_{k+1}}<n_{j_k}$. This 
would lead to a contradiction because every integer $n_i$ is positive. 
\nl Then there 
exist infinitely many integers $i$ such that  $\fa j\geq i$, $n_j\geq n_i$. The set of states 
is finite, hence there exists a state $q\in K$ and  infinitely many such 
integers $i_k$, $k\geq 1$, 
 such that for all $k\geq 1$, $q_{i_k}=q$ and $n_{i_k}>0$ or 
for all $k\geq 1$, $q_{i_k}=q$ and $n_{i_k}=0$ . Now if $Inf(r)=F$, the states not 
in $F$ occur only 
finitely many times during run $r$ thus  there exist two integers $s<s'$ such that 
the set of states \aut enters  between steps $i_s$ and $i_{s'}$ of the run $r$  is exactly $F$.

\begin{Rem}\label{rem}
The proof of Lemma \ref{loop-lemma} relies on a simple property of local minima of 
functions mapping natural numbers to themselves. A similar argument is 
due to Linna \cite{lin77}. 
\end{Rem}

\noi Then we shall write
\begin{enumerate}
\ite[(a)]
$(q, I) \stackrel{F}{\mapsto} ^\star (q, I^+)~~~~ \mbox{  if } n_{i_s}>0 \mbox{ and  }
 n_{i_{s'}}> n_{i_s}$
\ite[(b)]
$(q, I) \stackrel{F}{\mapsto}^\star (q, I^=)~~~~ \mbox{  if } n_{i_s}>0 \mbox{ and  }
 n_{i_{s'}}= n_{i_s}$
\ite[(c)]
$(q, Z_0) \stackrel{F}{\mapsto}^\star (q, Z_0)~~~~ \mbox{  if } n_{i_s}=0 \mbox{ and  }
n_{i_{s'}}=0$
\end{enumerate}

\noi The set $F$ is said to be an essential set (of states) and  we shall say that in the case 
$(a)$ there exists a loop $L(q, I, F, +)$, in the case $(b)$ 
there exists a loop $L(q, I, F, =)$, in the case $(c)$ 
there exists a loop $L(q, Z_0, F, =)$. Such a loop is positive if  $F \in \mathcal{F}$ and 
it is negative if $F \notin \mathcal{F}$. We then denote the loop $L(q, I, F, =)$ by
 $L^+(q, I, F, =)$ or  $L^-(q, I, F, =)$ and similarly in the other cases. 

\begin{lemma}
The set of essential sets and the set of positive and negative loops of a MBCA is effectively 
computable. 
\end{lemma}

\noi This follows  from the decidability of the emptiness problem 
for context free languages accepted by pushdown automata.

\hs 
We assume now some familiarity with the Wagner hierarchy as presented in \cite{wag} \cite{sta}.
The next step is to define, following Wagner's study, the (alternating) chains. 
Let $E^+$ (respectively $E^-$) be the set of  essential sets in $\mathcal{F}$ 
(respectively not in $\mathcal{F}$). An alternating chain of length $n$ is in the form 
$$F_1 \subset F_2 \subset F_3  \subset \ldots F_n$$
\noi where $F_i \in E^+$ iff $F_{i+1} \in E^-$ for $1\leq i<n$. It is a positive chain if 
$F_1\in E^+$ and a negative chain if $F_1\in E^-$. 

\hs As in the case of Muller automata \cite{sta}, 
one can see that if $F$ is a maximal essential set
then all (alternating) chains of maximal length contained in $F$ have the same sign (positive 
or negative) because in every chain of maximal length contained in $F$ one can replace 
the last essential set by $F$ itself. Let then $l(F)$ be the maximal length of chains contained 
in $F$ and $s(F)$ be the sign of these chains. 
\nl We now define the first invariant of the MBCA \aut as m(\aut) being the maximal length 
of chains of essential sets. Lemma \ref{loop-lemma} is crucial because it makes  
every essential set $F_i$ of a chain $F_1 \subset F_2 \subset F_3  \subset \ldots F_n$
to be indefinitely reachable from $(q, I)$ ( respectively $(q, Z_0)$)  if there 
exists a loop $L(q, I, F_n, +~or~=)$, ( respectively $L(q, Z_0, F_n, =)$). 

\hs The great difference between the case of Muller automata and the case of MBCA comes 
with the notion of superchain. Briefly speaking in a MA \aut a superchain of length $n$ is a 
sequence $S_1, \ldots ,S_n$ of chains of length m(\aut) such that for every integer 
$i$, $1\leq i <n$, $S_{i+1}$ is reachable from $S_i$ and $S_{i+1}$ is positive iff  
$S_i$ is negative. In the case of MA, $S_i$ cannot be reachable from $S_{i+1}$ otherwise 
there would exist a chain of length $>$m(\aut). 

\hs But in the case of MBCA, in such a superchain,  $S_i$ may be reachable from $S_{i+1}$ but 
{\bf with a reachability which is limited by the counter}. This leads to 
the notion of superchains of length $\om$, where $\om$ is the first infinite ordinal, 
and next of length $\alpha$ where $\alpha$ is an ordinal $<\om^2$. 

\hs An example of a MBCA \aut with m(\aut)$=m$ and a superchain of length $\om$ is 
obtained from two 
MA $\mathcal{B}$ and $\mathcal{B'}$ such that the graph of $\mathcal{B}$ is just constituted 
by a positive chain of length $m$ with a maximal essential set $F_m=\{q_1,\ldots q_m\}$ and 
the graph of $\mathcal{B'}$ is just constituted 
by a negative  chain of length $m$ with a maximal essential set $F'_m=\{q'_1,\ldots q'_m\}$.  
The behaviour of the MBCA \aut is as follows: at the beginning of an infinite run, the counter 
may be increased up to a counter value $N$; 
then there exist transitions from state $q_1$ to $q'_1$ 
and conversely from state $q'_1$ to $q_1$ but these transitions make the counter value 
decrease. Moreover 
\aut has also the transitions of the two MA $\mathcal{B}$ and $\mathcal{B'}$ but 
these transitions do not change the counter value.  Then one can see thet after a 
first transition from state $q_1$ to $q'_1$ or from $q'_1$ to $q_1$ 
the number of such transitions is bounded by the 
counter value $N$, but this initial value may be chosen $>n_0$ where $n_0$ is any given 
integer.

\hs Let then \aut be a  MBCA such that m(\aut)$=m$ and such that \aut has positive 
and negative chains of length $m$. A superchain of length $\om$ is formed by two maximal 
loops $L^+(q, I, F_m, +~or~=)$ and $L^-(q', I, F'_m, +~or~=)$ of such chains, i.e. 
$F_m$ is the last element of a  positive  chain of length $m$ and 
$F'_m$ is the last element of a   negative  chain of length $m$; moreover, for all $p_0>1$, 
configurations 
$(q, I^pZ_0)$ are reachable for integers $p>p_0$, and there exist 
transitions implying that 
$$(q, I^pZ_0) \mapsto^\star (q', I^{p'}Z_0) \mapsto^\star (q, I^{p''}Z_0)$$ 
\noi for some integers $p, p', p''$.  the MBCA \aut having not any chain of length 
$>m$, it holds that $p''<p$, because otherwise there would exist an essential 
set $F\supseteq F_m\cup F'_m$ and then there would exist a chain of length $>m$.
 And the loop $L^+(q, I, F_m, +~or~=)$ is in fact 
$L^+(q, I, F_m, =)$ and similarly $L^-(q', I, F'_m, +~or~=)$ is $L^-(q', I, F'_m, =)$ 
\nl One can informally say that $F_m$ is reachable from $F'_m$ and conversely 
but after such transitions the counter value has decreased hence there is a 
limitation to this reachability.

\begin{lemma}
The set of superchains of length $\om$ of a  MBCA is effectively 
computable. 
\end{lemma}

\noi Now one can define superchains of length $\om.p$ for an integer $p\geq 1$. 
Informally speaking a superchain of length $\om.p$ is a sequence 
$\Omega_1, \ldots , \Omega_p$ of superchains of length $\om$ such 
that  any  state $q$ of an essential set of $\Omega_{i+1}$ is reachable with unbounded 
values of the counter from 
any state of an essential set of $\Omega_{i}$.  
It is now easy to define superchains of length $\om.p + s\geq 1$, (with  $p, s$ some 
integers $\geq 0$), 
which are 
a sequence of a superchain of length $s$ followed by a superchain of length  $\om.p$. 
\nl In the case $s>0$, the superchain is said to be positive if it begins with a 
positive chain and it is said to be negative if it begins with a negative chain. 
\nl In the case $s=0$, we consider now that a superchain: 
$\Omega_1, \ldots , \Omega_p$, of length $\om.p$, 
is given with a loop $L$. Then it is said to be positive (respectively, negative) 
if $\Omega_1$ is formed by two maximal loops 
$L^+(q, I, F_m, =)$ and $L^-(q', I, F'_m, =)$ of  chains of length m(\aut)$=m$ and 
configurations $(q, I^pZ_0)$ are reachable for unbounded values of $p \geq 1$  from 
the  positive loop $L$ (respectively, from the negative loop $L$).

\hs We define now the second  invariant of the MBCA \aut as n(\aut) being the maximal length 
of superchains ( n(\aut) $< \om^2$ ). The MBCA is said to be prime if all 
superchains of length n(\aut) 
have the same sign, i.e. all are positive or all are negative. 
Denote s(\aut)$=0$ if \aut is not prime, s(\aut)$=1$ if 
all longest superchains are positive, and  s(\aut)$=-1$ if 
all longest superchains are negative. 

\begin{lemma} Let \aut be a MBCA. Then n(\aut) and  s(\aut) are computable. Moreover the 
set of superchains of length n(\aut) is computable. 
\end{lemma}

\noi We can now follow Wagner's study and define for $\alpha$ an ordinal $<\om^2$ and 
$m$ an integer $\geq 1$:

\hs $C_m^\alpha = \{ L({\bf \mathcal{A}})~/~$ s(\aut)$=1$ 
and m(\aut)$=m$ and n(\aut)=$\alpha \}$
\nl $D_m^\alpha = \{ L({\bf \mathcal{A}})~/~$ s(\aut)$=-1$ 
and m(\aut)$=m$ and n(\aut)=$\alpha \}$
\nl $E_m^\alpha = \{ L({\bf \mathcal{A}})~/~$ s(\aut)$=0$ 
and m(\aut)$=m$ and n(\aut)=$\alpha \}$

\hs Using the Wadge game, one can now show that each class $C_m^\alpha $ or 
$D_m^\alpha $ defines a Wadge degree, i.e. all \ol s in the same class 
$C_m^\alpha $ or $D_m^\alpha $ are Wadge equivalent. In other words 
$C_m^\alpha $ and $D_m^\alpha $ are the restrictions to the class {\bf BC} of some 
Wadge degrees. 
\nl Moreover when $\alpha=n$ is an integer, this degree corresponds to the degree 
obtained in the Wagner hierarchy for the classes $C_m^n $ or $D_m^n $. 

\hs The classes $C_m^\alpha$, $D_m^\alpha$, and $E_m^\alpha$, for $m$ an integer $\geq 1$ and 
$\alpha$ a non null ordinal $<\om^2$, form the coarse structure of the Wadge hierarchy 
of ${\bf BC}$. It is a strict extension of the coarse structure of the Wagner hierarchy 
studied in \cite{wag} and it  satisfies the following Theorem. 

\begin{theorem}
Let ${\bf \mathcal{A}}$ and ${\bf \mathcal{B}}$ be  two MBCA accepting the \ol s 
 $L({\bf \mathcal{A}})$ and $L({\bf \mathcal{B}})$. 
Then it holds that:
\begin{enumerate}
\ite  If m(${\bf \mathcal{A}})$ $<$ m(${\bf \mathcal{B}})$, then  
$L({\bf \mathcal{A}}) <_W L({\bf \mathcal{B}})$.  
\ite  If m(${\bf \mathcal{A}})$ $=$ m(${\bf \mathcal{B}})$, and  
 n(${\bf \mathcal{A}})$ $<$ n(${\bf \mathcal{B}})$, then 
$L({\bf \mathcal{A}}) <_W L({\bf \mathcal{B}})$.
\ite  If m(${\bf \mathcal{A}})$ $=$ m(${\bf \mathcal{B}})$, 
 n(${\bf \mathcal{A}})$ $=$ n(${\bf \mathcal{B}})$,  
s(${\bf \mathcal{A}})=1$ or s(${\bf \mathcal{A}})=-1$, and 
s(${\bf \mathcal{B}})=0$, then  $L({\bf \mathcal{A}}) <_W L({\bf \mathcal{B}})$.
\ite If m(${\bf \mathcal{A}})$ $=$ m(${\bf \mathcal{B}})$, 
 n(${\bf \mathcal{A}})$ $=$ n(${\bf \mathcal{B}})$,  
s(${\bf \mathcal{A}})=1$ and s(${\bf \mathcal{B}})=-1$, 
\nl then $L({\bf \mathcal{A}})$ and $L({\bf \mathcal{B}})$ are non self dual and 
$L({\bf \mathcal{A}}) \equiv_W L({\bf \mathcal{B}})^-$.
\end{enumerate}
\end{theorem}

\noi From this Theorem one can easily infer that the integer  m(${\bf \mathcal{A}})$,  the 
ordinal n(${\bf \mathcal{A}})$, and s(${\bf \mathcal{A}})\in \{-1, 0, 1\}$,  
are invariants of the \ol~
$L({\bf \mathcal{A}})$ and not only of the MBCA ${\bf \mathcal{A}}$:

\begin{corollary} Let  ${\bf \mathcal{A}}$ and ${\bf \mathcal{B}}$ be  two MBCA accepting the 
same \ol~  $L({\bf \mathcal{A}})=L({\bf \mathcal{B}})$. 
 Then m(${\bf \mathcal{A}})$ $=$ m(${\bf \mathcal{B}})$, 
n(${\bf \mathcal{A}})$ $=$ n(${\bf \mathcal{B}})$,  and 
s(${\bf \mathcal{A}})=$ s(${\bf \mathcal{B}})$.  

\end{corollary}

\noi One can give a canonical member in each of the classes 
$C_m^\alpha$, $D_m^\alpha$, and $E_m^\alpha$, for $m$ an integer $\geq 1$ and 
$\alpha$ a non null ordinal $<\om^2$.  And  one can easily deduce that the length 
of the  coarse structure of the Wadge hierarchy of blind counter \ol s is 
the ordinal $\om^3$, while 
the length of the coarse structure of the Wagner hierarchy was the ordinal $\om^2$. 

\hs The coarse structure of the class {\bf BC} is effective but it is not exactly the 
Wadge hierarchy of {\bf BC}, because  
 each  class $E_m^\alpha$ is the union of countably many (restrictions of) Wadge 
degrees. We can next  define a sort of derivation as Wagner did for Muller automata. 

\hs  Two MBCA  ${\bf \mathcal{A}}$ and ${\bf \mathcal{B}}$ in the same class $E_m^\alpha$ 
have essentially the same "most difficult parts" because they have 
positive and negative superchains of length 
n(${\bf \mathcal{A}})$ $=$ n(${\bf \mathcal{B}})$. 
Hence, in the case of Muller automata (then $\alpha$ is an integer), 
 Wagner's idea was to cut off the superchains of length 
n(${\bf \mathcal{A}})$ $=$ n(${\bf \mathcal{B}})$ of 
$\mathcal{A}$ and $\mathcal{B}$; this way one get some new automata 
$\partial {\bf \mathcal{A}}$ and $\partial {\bf \mathcal{B}}$ which are called 
the derivations of ${\bf \mathcal{A}}$ and ${\bf \mathcal{B}}$ and the comparison of 
${\bf \mathcal{A}}$ and ${\bf \mathcal{B}}$ with regard to 
$\leq_W$ is reduced to the comparison of 
their derivations $\partial {\bf \mathcal{A}}$ and $\partial {\bf \mathcal{B}}$.  

\hs In the case of MBCA one do as in the case of MA but with some modification. 
 We first define the derivation $\partial {\bf \mathcal{A}}$ of a MBCA in $E_m^\alpha$: 
 \aut$=(K,\Si,\Gamma, \delta, q_0, Z_0, \mathcal{F})$ as follows. 
\nl 
Let  $\partial K$ be the set of states in $K$ from which some positive 
{\bf and } some negative superchains of length n(${\bf \mathcal{A}})$ are reachable. 
In fact for each such $q\in \partial K$, it may exist an integer $n_q$ such that 
positive {\bf and }  negative superchains of length n(${\bf \mathcal{A}})$ are reachable
only from configurations $(q, I^nZ_0)$ with $n\geq n_q$. And these integers $n_q$ are 
effectively computable.  Let us define now 

$$\partial {\bf \mathcal{A}}=(\partial K,\Si,\Gamma=\{I, Z_0\}, \partial \delta, 
q_0, Z_0, \partial \mathcal{F})$$

\noi where  $\partial \delta$ is defined by: 

\hs  
for each $q\in \partial K$, $a\in \Si$,  $Z\in \Gamma$:
\nl  $\partial \delta(q, a, Z)=\delta(q, a, Z)$ if $\delta(q, a, Z)=(p, \gamma)$ for some 
$\gamma\in \Gas$ and $p\in \partial K$.  
\nl Otherwise $\partial \delta(q, a, Z)$ is undefined. 

\hs And $\partial \mathcal{F}=\{ F ~/~ F\subseteq \partial K \mbox{ and }  F\in \mathcal{F}\} $

\hs We consider now the MBCA $\partial {\bf \mathcal{A}}$ given with the integers $n_q$, for 
$q\in  \partial K$. Then we study the loops of  $\partial {\bf \mathcal{A}}$ as above 
but {\bf we keep only loops in the form
$L(q, Z_0 \mbox{ or } I, F, + or -)$  such that state $q$ is reachable 
with a counter value $n\geq n_q$}. We can next define chains and superchains 
for $\partial'{\bf \mathcal{A}}$=($\partial {\bf \mathcal{A}}, (n_q)_{q\in\partial K}$).  
We define m($\partial'{\bf \mathcal{A}})$, n($\partial'{\bf \mathcal{A}})$, 
and s($\partial'{\bf \mathcal{A}})$, and it holds that 
m($\partial'{\bf \mathcal{A}})$ $<$ m(${\bf \mathcal{A}})$. We then  
attribute a class  
$C_{\rm{m}(\partial'{\bf \mathcal{A}})}^{\rm{n}(\partial'{\bf \mathcal{A}})}$, 
$D_{\rm{m}(\partial'{\bf \mathcal{A}})}^{\rm{n}(\partial'{\bf \mathcal{A}})}$, 
or  $E_{\rm{m}(\partial'{\bf \mathcal{A}})}^{\rm{n}(\partial'{\bf \mathcal{A}})}$, 
 to $\partial'{\bf \mathcal{A}}$ as we did for ${\bf \mathcal{A}}$.  It may happen that 
there does not exist any loop for  $\partial'{\bf \mathcal{A}}$=($\partial {\bf \mathcal{A}}, 
(n_q)_{q\in\partial K}$); in that case we associate the class $E$ to 
$\partial'{\bf \mathcal{A}}$.  
Now we can iterate this process and associate to the MBCA ${\bf \mathcal{A}}$ a name 
$N({\bf \mathcal{A}})$ which is inductively defined by:

\begin{enumerate}
\ite If ${\bf \mathcal{A}}$ is prime and s(${\bf \mathcal{A}})=1$, then 
$N({\bf \mathcal{A}})=C_{\rm{m}({\bf \mathcal{A}})}^{\rm{n}({\bf \mathcal{A}})}$.
\ite  If ${\bf \mathcal{A}}$ is prime and s(${\bf \mathcal{A}})=-1$, then 
$N({\bf \mathcal{A}})=D_{\rm{m}({\bf \mathcal{A}})}^{\rm{n}({\bf \mathcal{A}})}$.
\ite If ${\bf \mathcal{A}}$ is not prime then 
$N({\bf \mathcal{A}})=E_{\rm{m}({\bf \mathcal{A}})}^{\rm{n}({\bf \mathcal{A}})}
N(\partial'{\bf \mathcal{A}})$. 
\end{enumerate}

\noi This name depends only on the \ol~ $L({\bf \mathcal{A}})$ accepted by the MBCA 
${\bf \mathcal{A}}$ and is effectively computable. 
We can write it in a similar fashion as in Wagner's study: we associate with each 
blind counter \ol~ L(\aut) in {\bf BC} a name in the form:

$$N(\mbox{\aut}) = E_{m_1}^{\alpha_1} \ldots E_{m_k}^{\alpha_k}H_{m_{k+1}}^{\alpha_{k+1}}$$
\noi where $m_1>m_2>\ldots >m_k>m_{k+1}$ are integers;  
each $\alpha_i$ is an ordinal $<\om^2$; and $H\in \{C, D\}$, or in the form:

$$N(\mbox{\aut}) = E_{m_1}^{\alpha_1} \ldots E_{m_k}^{\alpha_k}E$$
 
\noi which we shall simply denote by 
$$N(\mbox{\aut}) = E_{m_1}^{\alpha_1} \ldots E_{m_k}^{\alpha_k}$$

\noi where $m_1>m_2>\ldots >m_k$ are integers and   
each $\alpha_i$ is an ordinal $<\om^2$. 

\hs One can show that  each such name is really the name of an \ol~ in  {\bf BC}. 
And  the Wadge relation  $\leq_W$ is now computable because of the following result. 

\begin{theorem}
Let  ${\bf \mathcal{A}}$ and ${\bf \mathcal{B}}$ be  two MBCA accepting the \ol s 
 $L({\bf \mathcal{A}})$ and $L({\bf \mathcal{B}})$. Assume that the names associated 
with the MBCA ${\bf \mathcal{A}}$ and ${\bf \mathcal{B}}$ are:
$$N(\mbox{\aut}) = E_{m_1}^{\alpha_1} \ldots E_{m_k}^{\alpha_k}H_{m_{k+1}}^{\alpha_{k+1}}$$
$$N({\bf \mathcal{B}}) = 
E_{m'_1}^{\alpha'_1} \ldots E_{m'_l}^{\alpha'_l}H_{m'_{l+1}}^{'\alpha'_{l+1}}$$
\noi where  ($H=E$ or $H=C$ or $H=D$), and ($H'=E$ or $H'=C$ or $H'=D$). 
\nl Then $L({\bf \mathcal{A}}) \leq_W L({\bf \mathcal{B}})$ if there exists an integer 
$j\leq min(k+1, l+1)$ such that $m_i=m'_i$ and $n_i=n'_i$ for $1\leq i\leq j$ and one of the two
following properties holds.
\begin{enumerate}
\ite $j=k+1 \leq l+1$ and $H'=E$ or $H=H'$. 
\ite $j < min(k+1, l+1)$ and 
\nl $m_{j+1}< m'_{j+1}$ or ($m_{j+1}= m'_{j+1}$ and 
$\alpha_{j+1}< \alpha'_{j+1}$).  
\end{enumerate}
\end{theorem}

\noi Then the structure of the Wadge hierarchy of \ol s in {\bf BC} is completely determined. 
One can show that a blind counter \ol~ $L({\bf \mathcal{A}})$, where 
${\bf \mathcal{A}}$ is a MBCA, is in the class ${\bf \Delta^0_2 }$ iff 
m(${\bf \mathcal{A}})<2$, i.e. iff the name of ${\bf \mathcal{A}}$ is in the form 
$C_1^\alpha$, $D_1^\alpha$, or 
$E_1^\alpha$, for $\alpha < \om^2$. Thus the Wadge hierarchy restricted to the class 
{\bf BC}$\cap {\bf \Delta^0_2}$ has length $\om^2$, while the Wadge hierarchy 
restricted to $REG_\om \cap {\bf \Delta^0_2}$ has length $\om$. The Wadge hierarchy 
of {\bf BC}$\cap {\bf \Delta^0_2}$  is then a  great extension of the Wagner hierarchy
restricted to the class ${\bf \Delta^0_2}$. This phenomenon is still true for larger 
Wadge degrees and non  ${\bf \Delta^0_2}$-sets. Considering the length of the whole hierarchy 
of {\bf BC} we get the following:

\begin{corollary}
\begin{enumerate}
\ite[(a)] The length of the Wadge hierarchy of blind counter \ol s in  ${\bf \Delta^0_2}$
is $\om^2$.  
\ite[(b)] The length of the Wadge hierarchy of blind counter \ol s is the ordinal $\om^\om$ 
(hence it is equal to the length of the Wagner hierarchy).  
\end{enumerate}
\end{corollary}

\noi Once the structures of  two MBCA  ${\bf \mathcal{A}}$ and ${\bf \mathcal{B}}$ 
are  determined as well as their names $N(\mbox{\aut})$ and $N({\bf \mathcal{B}})$ 
are  effectively computed, one can construct winning strategies in Wadge games 
$W(L({\bf \mathcal{A}}), L({\bf \mathcal{B}}))$  and 
 $W(L({\bf \mathcal{B}}), L({\bf \mathcal{A}}))$. These strategies may be defined 
by blind counter transducers, and this extends Wagner's result to blind counter 
automata.   

\section{Concluding Remarks}

\noi This extended abstract is still a very summarized presentation of our results, 
which will need exposition of many other details we could not include in this paper \cite{aeff}. 

\hs  We have considered above \de real time blind counter  automata, which form a subclass 
of the class of \de pushdown automata and of the class of \de $k$-blind counter automata. 
The Wadge hierarchies of \ol s in each of these classes have been determined in a 
non effective way, by other methods, in \cite{dupcf} \cite{finb} \cite{wadpn}, and these 
 results had been announced in the survey \cite{dfr}. 
The Wadge degrees in these hierarchies may be described with similar names  
$$N(\mbox{\aut}) = E_{m_1}^{\alpha_1} \ldots E_{m_k}^{\alpha_k}H_{m_{k+1}}^{\alpha_{k+1}}$$
\noi where $m_1>m_2>\ldots >m_k>m_{k+1}$ are integers $\geq 1$ and  $H\in \{C, D, E\}$, and  

\begin{enumerate}
\ite each $\alpha_i$ is an ordinal $<\om^{k+1}$,  
in the case of {\bf $k$-blind counter automata.} 
\ite each $\alpha_i$ is an ordinal $<\om^\om$, 
 in the case of {\bf \de pushdown automata.} 
\end{enumerate}

\noi  We will further extend the results of the present paper in both directions 
to get decidability results and effective winning strategies in Wadge games. The above 
case of (one) blind counter automata  already introduces  some of the fundamental 
ideas which we will apply in further cases. 

\hs Another problem is to study  the complexity of the problem: 
" determine the Wadge degree of a blind counter \ol~", 
  extending this way the results of 
Wilke and Yoo to blind counter \ol s.  
\nl  Further study would be the investigation of links between the 
problems of simulation and bisimulation \cite{jan} \cite{jkm} \cite{jms} \cite{kuca} 
and the problem of finding winning strategies in Wadge games. 

\hs A Wadge game between two blind counter \ol s, whose complements are 
also blind  counter \ol s, can easily be reduced 
to a Gale-stewart game, (see \cite{tho95} \cite{pp}), with a winning set accepted by 
a \de  2-blind-counter automaton. This suggests that Walukiewicz's result, the proof of 
 the existence of effective winning strategies in a Gale-stewart 
game with a winning set accepted by a 
\de pushdown automaton, \cite{wal}, 
 could  be extended to the case of a winning set accepted by a \de 
multi  blind counter automata, giving additional results as asked by Thomas in 
\cite{tho95}. 

\hs {\bf \large Acknowledgements.} Thanks to  Jean-Pierre Ressayre and Jacques Duparc
for many helpful discussions about Wadge and Wagner Hierarchies.
\nl Thanks also  to the anonymous referees
for useful comments on the preliminary version of this paper. In particular the remark 
\ref{rem} is due to one of them.

\end{document}